\documentstyle[12pt,dina4,worldsci]{article}
\begin{document}
\hfuzz = 2.5pt
\bibliographystyle{unsrt}
\newcommand{\al}{\alpha}
\newcommand{\be}{\beta}
\newcommand{\ga}{\gamma}
\newcommand{\hodge}{*}
\newcommand{\contr}{\hskip.7mm
                    \rule[.1mm]{2.3mm}{.2mm}\rule[.1mm]{.2mm}{2.mm}\hskip.7mm}
\newcommand{\fat}[1]{{\bf #1}}
\newcommand{\beq}{\begin{equation}}
\newcommand{\eeq}{\end{equation}}
\newcommand{\beqq}{$$}
\newcommand{\eeqq}{$$}
\newcommand{\p}{\partial}
\newcommand{\pq}[2]{\frac{\p#1}{\p#2}}
\newcommand{\ce}{$\,\!$}
\title{THE USE OF COMPUTER ALGEBRA IN MAXWELL'S THEORY\thanks{Supported by
                           {\em the Graduiertenkolleg `Scientific Computing'},
               K\"oln--St.Augustin.}\thanks{Computer Algebra
                                  in Science and Engineering. Proc.\ of the
      ZiF-Workshop, Bielefeld, Aug.\ 1994. J. Fleischer et al.\ (eds.) World
      Scientific, Singapore, to be published (1995).}
}
\author{ROLAND A. PUNTIGAM\\
           {\em Institute for Theoretical Physics, University of Cologne\\
                D--50923 K\"oln, Germany}\\
                \vspace{0.3cm}
       EBERHARD SCHR\"UFER\\
           {\em German National Center for Computer Science (GMD)\\
                D--53757 St.Augustin, Germany}\\
                \vspace{0.3cm}
       and \\
                \vspace{0.3cm}
       FRIEDRICH W. HEHL\\
           {\em Institute for Theoretical Physics, University of Cologne\\
                D--50923 K\"oln, Germany}
}
\maketitle
\begin{abstract}
We present a small computer algebra program for use in Maxwell's
theory. The Maxwell equations and the energy-momentum current of the
electromagnetic field are formulated in the language of exterior
differential forms. The corresponding program can be applied (in the
presence of a gravitational field) in curved Riemannian spacetime as
well as in flat Minkowski spacetime in inertial or {\em non-inertial}
frames. Our program is written for the computer algebra system REDUCE
with the help of the EXCALC package for exterior differential forms.\par

The two major advantages of this modern approch to electrodynamics ---
the natural formulation, free of both metric and coordinates, and the
straightforward programming of problems  --- are illustrated by examining a
number of examples ranging from the Coulomb field of a static point charge to
the Kerr--Newman solution of the Einstein--Maxwell equations.
\end{abstract}
\section{Introduction}
In the lecture of Hartley\cite{Hartley94}\ce, the effective use of exterior
differential forms has been demonstrated convincingly.
We were led to our Maxwell program because we were searching for electrically
charged solutions in the framework of general relativity --- see the talk of
Brans \cite{Brans94} in this context --- and of alternative gravitational
theories\cite{Baekler88a}\ce, and only streamlining our program using
exterior differential forms made it a feasible task.
We believe that our program, besides being helpful in physics, could also
have applications in electrical
engineering (see the textbook of Meetz and Engl\cite{Meetz80}\ce, written
for engineering students, inter alia, where the calculus of exterior forms is
applied throughout).\par

Our axiomatic formulation of Maxwell's theory
is based on {\em three} principles standing on firm experimental
bases: (i) the conservation of electric charge, (ii) the notion of the Lorentz
force, and (iii) the conservation of electromagnetic flux.
\begin{figure}
\vspace{8 truecm}
\caption{Graphic representation of geometric objects used in Maxwell's theory
         (adapted from Burke\protect\cite{Burke83,Burke85} and
         Konzen\protect\cite{Konzen94}).\label{edobjects}}
\end{figure}
\section{Maxwell's Theory in Brief}
\subsection{Electric Charge Conservation}
The inhomogeneous set of Maxwell's equations, in terms of exterior differential
forms, may be written as
  \beq\label{InhomMax3}
      \left\{ \begin{array}{r@{\quad=\quad}l}
          \underline{d}\,{\cal H} - \dot{\cal D} & j\, , \\
          \underline{d}\,{\cal D}                & \rho\, .
      \end{array}\right\}
  \eeq
The sources are the electric current density 2--form $j$ and the electric
charge density 3--form $\rho$.
The magnetic excitation 1--form is denoted by ${\cal H}$ and the electric
excitation 2--form by ${\cal D}$. The underline $\underline{d}$ is the
3--dimensional exterior derivative, and the dot denotes the time
derivative.  \par

These equations, reformulated in 4 dimensions, read
  \beq\label{InhomMax}
     d H = J\, ,
  \eeq
provided we introduce the time-space decompositions of the
electromagnetic excitation 2--form ($t =$ time coordinate)
  \beq\label{H4dim}
     H = -{\cal H}\wedge dt +{\cal D}
  \eeq
and the electric current 3--form
  \beq
     J = -j\wedge dt +\rho\, .
  \eeq
If the spacetime under consideration is contractable,
then it can be shown that (\ref{InhomMax}) can be derived from the
integral version
  \beq\label{ConservCharge}
     \oint_{\partial V_4}J = \int_{V_4} dJ = 0
  \eeq
of the conservation law for electric
charge\cite{Truesdell60,Post62,Post82}\ce.

\subsection{Lorentz Force}\label{Sec:LorentzForce}
Postulating the (vectorial) Lorentz-force equation as\footnote{In 4
dimensions, instead of (\ref{LorentzForce}), we have
$f_\al = (e_\al \contr F) \wedge J$, with a 4--dimensional frame
$e_\al = \{e_0,e_1,e_2,e_3\}$.}
  \beq\label{LorentzForce}
     \fat f = \rho \wedge \fat E + j \wedge \fat B\, ,
  \eeq
we are led in a most natural way to the electrical field strength 1--form
$E$ and the magnetic field strength 2--form $B$, with
$\fat E = \fat e \contr E$ and $\fat B = \fat e \contr B$, where
$\fat e = \{e_1,e_2,e_3\}$ represents an arbitrary 3-dimensional vector basis
(frame), a so--called triad, and $\contr$ denotes the interior product.  In
analogy to (\ref{H4dim}), the field strength can be collected in the
4-dimensional electromagnetic field strength 2--form
  \beq\label{4DFStrength}
     F = -dt \wedge E + B\, .
  \eeq
One possible graphic representation of the sources and fields of Maxwell's
theory is displayed in Fig.\ref{edobjects}. We find this way of representing
exterior differential forms quite intuitive and, for more details, refer to
the literature cited in the caption of Fig.\ref{edobjects}.
\subsection{Electromagnetic Flux Conservation}
The homogeneous set of Maxwell's equations
  \beq\label{HomMax3}
      \left\{ \begin{array}{r@{\quad=\quad}l}
          \underline{d} E  + \dot B & 0\, , \\
          \underline{d} B           & 0\, ,
      \end{array}\right\}
  \eeq
can be rewritten by means of (\ref{4DFStrength}) as
  \beq\label{HomMax}
     d F = 0\, ,
  \eeq
or, in integral form, as an electromagnetic flux conservation law
  \beq\label{ConservFlux}
     \oint_{\partial V_3} F = \int_{V_3} dF = 0\, .
  \eeq

\subsection{Maxwellian Structure and Spacetime}
So far we have the two Maxwell equations (\ref{InhomMax}) and (\ref{HomMax})
and the expression for the Lorentz force (\ref{LorentzForce})
interrelating the current and
field strength by means of the force concept of mechanics. These
equations can be formulated on any 4-dimensional differential
manifold, as long as it allows a (1+3)--decomposition.
It is worth noting that these Maxwellian structures emerge prior to the
introduction of a metric or a connection.
The conservation laws (\ref{ConservCharge}) and
(\ref{ConservFlux}) only require the
possiblity to circumscribe a boundary around an arbitarary 4- or
3-dimensional volume element, and to {\it count} the
charge or flux units contained therein.

In particular this means that the equations (\ref{InhomMax}), (\ref{HomMax}),
and (\ref{LorentzForce}) are valid in this form in a curved and contorted
spacetime, i.e. with
non-vanishing curvature and torsion, or in a flat Minkowskian
spacetime in inertial {\em and} non-inertial frames -- they always keep this
form.\par

However, in (\ref{InhomMax}) and (\ref{HomMax}) we have only 6
independent evolution equations, namely (\ref{InhomMax3})$_1$ and
(\ref{HomMax3})$_1$, for specifying the $2 \times 6$
independent fields $\{H,F\}$. Something is still missing, namely the
{\em constitutive laws}, such as the one relating $H$ to $F$, and
Ohm's law.

In a vacuum, the field strength and excitation are related by
\begin{equation}\label{Constit}
   H={1\over \mu}\,^\hodge F\,,\qquad{\rm with}\qquad
   \mbox{dim }\mu = \frac{action}{charge^2} = \frac{V}{A} = \Omega \, ,
\end{equation}
where $\hodge$ is the Hodge star operator, which does depend on the metric
--- or rather on its determinant.

\section{Program}\label{Program}
In an interactive Reduce 3.5 session\cite{Hearn93,Hehl93,Stauffer93}\ce,
the Excalc package\cite{Schrufer94} can be loaded by the
command\footnote{Soleng\cite{Soleng94} has written the Mathematica package
CARTAN which should be also useful for programming Maxwell's theory.}
\begin{verbatim}
    load_package excalc$
\end{verbatim}\par

First Excalc has to be acquainted with the rank of the differential forms
to be used in the code. This is achieved by the declaration
\begin{verbatim}
    pform pot1=1, {farad2,excit2}=2, {maxhom3,maxinh3}=3$
\end{verbatim}
where {\tt pot1} and {\tt farad2} are the identifiers of the
electromagnetic potential and field strength, {\tt excit2} represents the
electromagnetic excitation, and {\tt maxhom3} and {\tt maxinh3} denote the
left hand sides of the homogeneous and the inhomogeneous Maxwell equations.
The numbers after the equal sign specify the rank of the forms.\par

The mathematical symbols used in exterior calculus, translate into Excalc
as listed in table \ref{ExcalcSymbols}.
\par
\begin{table}[htb]\begin{center}
\begin{tabular}{| c | c | c | c|}
\hline
Math. Symbol & Excalc Symbol    & Operator          & Operator Type\\[.1truecm]
\hline
$\wedge$   & \verb|^|  & exterior product      & ``nary'' infix    \\[.1truecm]
$d$        & {\tt d }  & exterior derivative   & unary prefix      \\[.1truecm]
$\partial$ & {\tt @ }  & partial derivative    & ``nary'' prefix   \\[.1truecm]
\contr     & {\tt \_|} & interior product      & binary infix      \\[.1truecm]
\pounds    & {\tt |\_} & Lie derivative        & binary infix      \\[.1truecm]
 $\hodge$   & {\tt \# } & Hodge star operator   & unary prefix
\\[.1truecm]
\hline
\end{tabular}\end{center}
\caption{\label{ExcalcSymbols}Translation of mathematical symbols into Excalc.}
\end{table}
Now we are ready to write a small prototype program. Suppose we
consider Minkowski spacetime and use polar coordinates
$\{x^0,x^1,x^2,x^3\} = \{t,r,\theta,\varphi\}$. Then we have
to specify a local basis of 1--forms, a coframe,
by means of the Excalc declaration
\begin{verbatim}
    coframe   o(t)     =                  d t,
              o(r)     =                  d r,
              o(theta) = r              * d theta,
              o(phi)   = r * sin(theta) * d phi
       with signature (1,-1,-1,-1);
    frame e;
\end{verbatim}
The command {\tt frame e;} provides us with the frame $e_\al$ dual to
the coframe that we have chosen to call $o^\al$.\par
\subsection{Static Point Charge}\label{Coulomb}
As test scenario, let us take the Coulomb field of a static point charge $q$.
The corresponding electromagnetic potential 1--form $A$ and the field strength
2--form $F = d A$ in Excalc read
\begin{verbatim}
    pot1   := -(q/r) * d t;
    farad2 := d pot1;
\end{verbatim}
Now we translate into Excalc the homogeneous Maxwell equation (\ref{HomMax})
\begin{verbatim}
    maxhom3 := d farad2;
\end{verbatim}
the constitutive law for vacuum (\ref{Constit})
\begin{verbatim}
    excit2  := (1/mu) * # farad2;
\end{verbatim}
and the inhomogeneous Maxwell equation (\ref{InhomMax})
\begin{verbatim}
    maxinh3 := d excit2;
\end{verbatim}
The result of these commands, typed in during an interactive
Reduce session, is that both left hand sides of Maxwell's equations are zero;
in other words, Maxwell's equations are fulfilled, and the ansatz for $A$ is
justified.\par

The canonical energy-momentum 3--form of Maxwell's field is given
by\footnote{The conventions we use for energy-momentum are as follows:
define the  energy-momentum tensor $T_{\beta\alpha}$ in terms the
energy-momentum 3--form
$\Sigma_\alpha$ by $\Sigma_\alpha = T_{\beta\alpha} \eta^\beta$, with
$\eta^\beta :={} ^\hodge o^\beta$.
If this $T_{\beta\alpha}$ is multiplied by $-1$, we have the conventions
of Ref.16. This  is the reason for the minus sign
on the right hand side of (\ref{EinsteinMaxwell}) below.}
\beq\label{MaxEnergy}
   \Sigma_\alpha = e_\alpha \contr L + (e_\alpha \contr F) \wedge H\,,
\eeq
where $L$ is the Lagrange 4--form of Maxwell's field:
\beq
   L= -{1\over 2}\, F\wedge H\,.
\eeq
We declare the corresponding forms
\begin{verbatim}
    pform lmax4=4, maxenergy3(a)=3$
\end{verbatim}
and define the Lagrangian
\begin{verbatim}
    lmax4 := -(1/2) * farad2 ^ excit2;
\end{verbatim}
Now the energy-momentum 3--form (\ref{MaxEnergy}) can be computed
straightforwardly
(note that a lower index is always marked with a minus sign in Excalc):
\begin{verbatim}
    maxenergy3(-a) := e(-a) _| lmax4 + (e(-a) _| farad2) ^ excit2;
\end{verbatim}
We find
  \beq
     \Sigma_t = - \frac{q^2}{2\mu r^4}\, o^r \wedge o^\theta \wedge o^\varphi
  \eeq
and (up to minus signs depending on the ordering of the basis 1--forms
$o^\al$) equivalent expressions for the three other components. As
expected, energy-momentum turns out to be proportional to $q^2/r^4$.
\subsection{Small Ring Current}\label{Dipole}
\begin{figure}
\vspace{8 truecm}
\caption{\label{DipoleField}The electromagnetic field of a small ring current.}
\end{figure}
An equally simple task is the potential of the magnetic dipole moment
$\vec{\cal M}$ of a small ring current. From the literature (see
Jackson\cite{Jackson75}\ce, p.182) we take the vector potential in the
Minkowski spacetime as $\vec A(\vec r) = \vec{\cal M} \times \vec r /r^3$. We
evaluate the vector product (see Fig. \ref{DipoleField}) and find
$\vec A = {\cal M}/r^2 \sin\theta\, \vec e_\varphi$, where $\vec e_\varphi$
is the unit vector in $\varphi$--direction. Then the four-potential 1--form
reads
\beq\label{DipolePot}
   A = \frac{{\cal M}}{r^2} \sin\theta\, o^\varphi =
       \frac{{\cal M}}{r} \sin^2\theta\, d \varphi.
\eeq\par

Let us denote the angular momentum of the electrically charged circling
particles by $L$ and the
angular momentum per unit mass by $a:= L/m$. Then the gyromagnetic parallelism
yields (Jackson\cite{Jackson75}\ce, p.183) ${\cal M} = g\,q\, a /2 $
with $q =$ charge or, substituted in (\ref{DipolePot}),
\beq\label{DipolePotB}
   A = g\,\frac{q\, a}{2 r}\sin^2\theta\, d \varphi.
\eeq
By $g$ we denote the $g$-factor of the current distribution, which equals 1
for orbital angular momentum (as for a ring current) and 2 for spin angular
momentum.\par

In Excalc we have
\begin{verbatim}
    pot1 := g*q*a/(2*r) * sin(theta)**2 * d phi;
\end{verbatim}
and, for field strength and excitation,
\begin{verbatim}
    farad2  := d pot1;
    maxhom3 := d farad2;
    excit2  := (1/mu) * # farad2;
    maxinh3 := d excit2;
\end{verbatim}
The Maxwell equations turn out to be fulfilled. We will apply the potential
(\ref{DipolePotB}) in Sec.\ref{KerrNewman} in a less trivial context.
\section{Examples}
\subsection{Electromagnetism in a Rotating Frame}\label{RotatingFrame}
In the flat spacetime of special relativity, {\it inertial} frames of
reference are realized by (pseudo-)orthonormal tetrads which are all
arranged in a parallel fashion --- here parallel is meant in the
ordinary (pseudo-)Euclidean way. A corresponding coframe, in polar
coordinates, was specified in Sec.\ref{Program}. If we want to describe
non-inertial frames of reference, then we can turn to an accelerated
and rotating frame, as used in Ref.18, for example, (see also
Ref.19), and the parallel arrangement of the frames will
be disturbed thereby.\par

There is a plethora of different books and papers on electrodynamics
referred to non-inertial frames, see Refs.20,21 and literature given
there.  Formulated in terms of exterior differential forms, such
formalisms simplify appreciably.  We will demonstrate this claim by
investigating the question of how a rotating observer will see the
field of the magnetic dipole as specified in Sec.\ref{Dipole}.\par

Let us assume that the axis of rotation coincides with the $z$-axis such that
$\omega$ is the only nonvanishing component of the local angular velocity. We
use the coframe\footnote{By using Excalc, one can prove easily
that the Riemannian curvature belonging to our new coframe vanishes
identically, that is, we are still in a flat Minkowski space.} as given
in Thirring\cite{Thirring93} (last equation on p.556)
in order to express this configuration in Excalc\footnote{Before making a
new {\tt coframe} statement, we recommend starting a new Reduce session.}:
\begin{verbatim}
    coframe
        o(t)   = (d t - omega*rho**2* d phi)/sqrt(1 - (omega*rho)**2),
        o(rho) = d rho,
        o(phi) = (d phi - omega* d t)*rho/sqrt(1 - (omega*rho)**2),
        o(z)   = d z
      with signature 1,-1,-1,-1;
    frame e;
\end{verbatim}
The dipole potential (\ref{DipolePotB}) is expressed in terms of polar
coordinates. To put it into our rotating cylindrical coordinates, we would
need a pullback operator for forms which will, however, only be available in
the next Excalc release. We shall overcome this deficiency by a small trick.
Define
\begin{verbatim}
    pform {r,theta}=0$

    r     := sqrt(rho**2 + z**2);
    theta := acos(z/sqrt(rho**2 + z**2));
\end{verbatim}
If we start with polar coordinates $\{t,r,\theta,\varphi\}$, Excalc
will then substitute cylindrical coordinates $\{t,\rho,\varphi,z\}$
instead. Using this transformation, the dipole potential
(\ref{DipolePotB}) can be viewed from the rotating system of
reference:
\begin{verbatim}
    factor o(0),o(1),o(2),o(3), ^$ off exp$ on gcd$
    pform pot1=1, {farad2,excit2}=2, {maxhom3,maxinh3}=3$

    pot1   := g*q*a/(2*r) * sin(theta)**2 * d phi;
    farad2 := d pot1;
\end{verbatim}
By making use of the {\tt factor} command we caused Reduce to ``factor out''
the terms given. Additionally, we turned off and on
the two Reduce switches {\tt exp} ({\em expand}) and {\tt gcd}
({\em greatest common divisor}), respectively, in order to simplify the
resulting expressions.\par

Let us have a closer look at the physical situation as seen by the rotating
observer. For this purpose, we compute the components of the electric and the
magnetic field, respectively (see Ref.19 for the
$(3+1)$--decomposition of the electromagnetic field strength 2--form):
\begin{verbatim}
    indexrange {i,j,k}={rho,phi,z}$
    pform {ee(i),bb(i,j)}=0$
    index_symmetries bb(i,j):antisymmetric$

    ee(-i)    := e(-i) _| (e(-t) _| farad2);
    bb(-i,-j) := e(-i) _| (e(-j) _| farad2);
\end{verbatim}
The {\tt indexrange} declaration allows to have certain indices not
ranging over all spacetime dimensions but over a subset that can be
chosen arbitrarily. Here we declared the indices $i, j$, and $k$ to
cover only the {\em spatial} section of spacetime.\par

The rotating observer still perceives, up to a Lorentz factor
$\gamma = 1/\sqrt{1-(\omega\rho)^2}$, the magnetic dipole field
\begin{eqnarray*}
    B_{\phi z} & = & - \frac{\gamma\;a g q}{2}
                  \frac{3\rho z}{(\rho^2 +z^2)^{5/2}}
             = - \gamma\;{\cal M} \frac{3\rho z}{(\rho^2 +z^2)^{5/2}}\\
    B_{z\rho} & = &  0\\
    B_{\rho\phi}    & = & \frac{\gamma\; a g q}{2}
                  \frac{(\rho^2 - 2z^2)}{(\rho^2 +z^2)^{5/2}}
             =  \gamma\;{\cal M}\frac{(\rho^2 - 2z^2)}{(\rho^2 +z^2)^{5/2}}
\end{eqnarray*}
(we substituted again the magnetic dipole moment ${\cal M} = a g q/2$), but in
addition he observes an {\em electric} field
\begin{eqnarray*}
    E_\rho & = & \frac{\gamma\;a g q}{2}
                 \frac{\omega \rho\; (\rho^2 - 2z^2)}{(\rho^2 +z^2)^{5/2}}
             =   \gamma\;{\cal M}
                 \frac{\omega \rho\; (\rho^2 - 2z^2)}{(\rho^2 +z^2)^{5/2}}\\
    E_\phi & = & 0\\
    E_z    & = & \frac{\gamma\;a g q}{2}
                 \frac{3 \omega \rho^2 z}{(\rho^2 +z^2)^{5/2}}
             = \gamma\;{\cal M}\frac{3 \omega \rho^2 z}{(\rho^2 +z^2)^{5/2}}
\end{eqnarray*}
whose components correspond to those of the magnetic field rotated by $\pi/2$
around an axis perpendicular to the $z$--axis and multiplied by
$\omega\;\rho$.\par

For completeness, we verify that the inhomogeneous Maxwell equation is still
satisfied:
\begin{verbatim}
    excit2  := (1/mu) * # farad2;
    maxinh3 := d excit2;
\end{verbatim}
\subsection{Deriving the Reissner--Nordstr\"om from the Schwarz\-schild
            Solution}\label{ReissnerNordstrom}
To give a less trivial example, we will derive the spacetime of a spherically
symmetric mass distribution carrying a net electric charge $q$. In general
relativity, the corresponding solution of the Einstein equation is called
the Reissner--Nordstr\"om solution.\par

We start from the {\em uncharged} spherically symmetric solution (the
Schwarzschild
solution). Its spacetime geometry is determined by the coframe that can
immediately be expressed in Excalc:
\begin{verbatim}
    pform psi=0$ fdomain psi=psi(r)$
    coframe o(t)     = psi            * d t,
            o(r)     = (1/psi)        * d r,
            o(theta) = r              * d theta,
            o(phi)   = r * sin(theta) * d phi
       with signature (1,-1,-1,-1)$
    frame e$
\end{verbatim}
Note that functions have to be declared as 0--forms also.
The {\tt fdomain} declaration specifies the independent variable(s) of a given
form. The Schwarzschild function $\psi$ has the form
$\psi = \sqrt{1 - 2m/r}$,
where $m$ is the mass of the object. It can be shown, see below,
that this orthonormal coframe satisfies the vacuum Einstein equation indeed.
\par

Now we want to generalize the Schwarzschild solution to a corresponding
electrically charged solution. We proceed in exactly the
same manner as in Sec.\ref{Coulomb}: the Schwarzschild coframe, with arbitrary
$\psi$, is already prescribed (and has been checked),
we type in (or read from a corresponding file) the commands for the point
charge $q$:
\begin{verbatim}
    pform pot1=1, {farad2,excit2}=2, {maxhom3,maxinh3}=3$

    pot1    := -(q/r) * d t;
    farad2  := d pot1;
    maxhom3 := d farad2;
    excit2  := (1/mu) * # farad2;
    maxinh3 := d excit2;
\end{verbatim}
We find that Maxwell's equations are still fulfilled, even though we did not
even specify the function $\psi$! Note that this only
turns out to be true since we specified the potential {\tt pot1} in
terms of $dt$ and not $o(t)$. However, the electromagnetic
energy--momentum 3--form $\Sigma_\al$ given in (\ref{MaxEnergy}) acts as a
source in the Einstein equation:
\beq\label{EinsteinMaxwell}
   G_\alpha = -\kappa \, \Sigma_\alpha \,.
\eeq
The Einstein 3--form $G_\alpha$ (called
{\tt einstein3} in the code) is defined according to
(see Ref.23, e.g.):
\beq
   G_\alpha := (1/2)\; R^{\beta\gamma} \wedge\ ^\hodge (o_\alpha\wedge
                o_\beta \wedge  o_\gamma) \,.
\eeq
Basically we had put a point charge on a prescribed
Schwarzschild geometry --- and, up to now, we forgot the gravitating character
of the electromagnetic field. The obvious consequence is that the corresponding
Einstein-Maxwell equation is {\em not} satisfied. In order to fix this, we
have to ``relax'' the Schwarzschild spacetime;
this is done by reformulating the coframe.
Since the charged solution sought should remain spherically symmetric,
we try to change the function $\psi$ alone. The most straightforward
ansatz that comes to mind is
\begin{verbatim}
    pform unknown=0$ fdomain unknown=unknown(r)$
    psi := sqrt(1 - 2 * m/r + unknown);
\end{verbatim}
The new function {\tt unknown} must be determined from the Einstein-Maxwell
equation (\ref{EinsteinMaxwell}).

We first have to calculate the Riemann or Levi--Civita connection
1--form $\Gamma^{\alpha\beta}$ and the curvature 2--form $R^{\beta\gamma}$
(with identifiers {\tt chris1} and {\tt curv2}). The corresponding Excalc
code reads\footnote{Note that the old {\tt antisymmetric <identifier>;}
declaration is unsupported already in Reduce 3.5. It has to be replaced by
{\tt index\_symmetries <identifier>(a,b):antisymmetric;}}
\footnote{The transposition according to {\tt chris1(a,b) :=
christ1(b,a)\$} is necessary to
convert the Excalc conventions for the connection 1--form into our own
convention, see Ref.24.}
\begin{verbatim}
    pform chris1(a,b)=1, curv2(a,b)=2, einstein3(a)=3$
    index_symmetries chris1(a,b), curv2(a,b):antisymmetric$
    riemannconx christ1$ chris1(a,b) := christ1(b,a)$

    curv2(a,b)    := d chris1(a,b) + chris1(-c,b) ^ chris1(a,c);
    einstein3(-a) := (1/2) * curv2(b,c) ^ #(o(-a)^o(-b)^o(-c));
\end{verbatim}
The Excalc command {\tt riemannconx <identifier>\$} initiates
the calculation of the connection 1--form and stores it in {\tt identifier}.
If {\tt unknown} vanishes, we have $G_\alpha = 0$, which
proves our earlier claim that the original coframe satisfies the Einstein
vacuum equation.\par

Subsequently, we compute the Maxwell energy--momentum, see (\ref{MaxEnergy}):
\begin{verbatim}
    pform lmax4=4, maxenergy3(a)=3$

    lmax4 := -(1/2) * farad2 ^ excit2;
    maxenergy3(-a) := e(-a) _| lmax4 + (e(-a) _| farad2)^excit2;
\end{verbatim}
If the vacuum Einstein-Maxwell equation is to be satisfied, the sum of
the Einstein 3--form and (a constant times) the Maxwell
energy--momentum 3--form must vanish:
\begin{verbatim}
    pform ode3(a)=3$
    ode3(a) := einstein3(a) + kappa * maxenergy3(a)
\end{verbatim}
A typical component of the output of {\tt ode3} reads
\beq
   ode3^\varphi =  \left(- \frac{1}{2}\pq{\mbox{\tt\,unknown}} {r^2}
                      - \frac{1}{r}\pq{\mbox{\tt\,unknown}} {r}
                      + \frac{\kappa}{2\mu}\frac{q^2}{r^4}\right)
             o^t \wedge o^r \wedge o^\theta.
\eeq
Requiring the coefficients to vanish yields a second order differential
equation for the function {\tt unknown}. It can be solved by means of the
ansatz {\tt unknown := a * r**n;} A short Reduce calculation
yields\footnote{If ode3 was of a more complicated nature, we could feed it
into Maple or Mathematica, which have at the moment a more complete ODE solver
than Reduce.}
\begin{verbatim}
    unknown := kappa/(2*mu) * (q/r)**2;
\end{verbatim}
This reproduces, when $\psi$ is evaluated, the well known Reissner--Nordstr\"om
function:
\begin{verbatim}
    psi := psi;
\end{verbatim}
\beq
    \psi := \sqrt{1 - 2 \frac{m}{r}
           +  \frac{\kappa}{2\mu} \left(\frac{q}{r}\right)^2}.
\eeq
It is remarkable that,
although we are now working on a curved spacetime mani\-fold, {\em not a
single command} of the Maxwell code had to be changed. This should clearly
demonstrate the simplifying value of the formulation of Maxwell's theory
presented here.
\subsection{Deriving the Kerr-Newman from the Kerr Solution}\label{KerrNewman}
In analogy to Sec.\ref{ReissnerNordstrom}, we will determine the solution to
the Einstein-Maxwell equation of a {\em rotating} axial symmetric
mass distribution with net charge $q$. It is called the Kerr-Newman solution.

Again we take as starting point the spacetime of the corresponding
{\em uncharged} solution, the Kerr solution. Using
Boyer-Lindquist coordinates
$\{t,\rho,\theta,\varphi\}$, the orthonormal coframe reads:
\begin{verbatim}
    pform {rr,delsqrt}=0$
    fdomain rr=rr(rho,theta),delsqrt=delsqrt(rho)$
    coframe
      o(t)     = (delsqrt/rr)      *(d t - (a*sin(theta)**2) * d phi),
      o(rho)   = (rr/delsqrt)      * d rho,
      o(theta) = rr                * d theta,
      o(phi)   = 1/rr * sin(theta) *(a*d t - (rho**2 + a**2) * d phi)
    with signature 1,-1,-1,-1$
    frame e$
\end{verbatim}
The functions {\tt delsqrt} and {\tt rr} depend on the parameters $m$ (mass)
and $a$ (angular momentum per unit mass). They have the form
${\tt delsqrt} = \sqrt{\rho^2 + a^2 - 2\,m\,\rho}$ and
${\tt rr} = \sqrt{\rho^2 + (a \, \cos\theta)^2}$.
We can prove (in the same way as in Sec.\ref{ReissnerNordstrom}) that this
coframe satisfies the vacuum Einstein equation.\par

To determine the Kerr-Newman solution, we first have to specify the
electromagnetic potential. To start with, we take the potential of a rotating
charge distribution which we approximate to lowest
order of $1/\rho$ by a static point charge and a magnetic dipole (see
Secs.\ref{Coulomb} and \ref{Dipole}):
\begin{verbatim}
    pform pot1=1, {farad2,excit2}=2, {maxhom3,maxinh3}=3$
    pot1 := -(q/rho) * (d t - g/2*a*sin(theta)**2 * d phi);
\end{verbatim}
Here the $g$-factor is to be determined.\par

In Minkowski spacetime, the homogeneous and inhomogeneous Maxwell equations
are satisfied, as we have seen in Secs.\ref{Coulomb} and \ref{Dipole}
(a superposition of solutions is again a solution).
However, now we have to compute Maxwell's equations on the curved Kerr
spacetime:
\begin{verbatim}
    let {(sin ~x)**2 => 1 - (cos x)**2}$ factor ^$
    farad2  := d pot1;
    maxhom3 := d farad2;
    excit2  := (1/mu) * # farad2;
    maxinh3 := d excit2;
\end{verbatim}
We find that the Hodge star operator spoils (by means of $\det g$)
the inhomogeneous Maxwell equation.
Since we expect that this failure is connected to the $1/\rho$ behavior of the
potential, we change the potential by the following ansatz
\begin{verbatim}
    pform f=0$ fdomain f = f(rho,theta)$
    pot1 := -q/(rho * f) * (d t - g/2*a*sin(theta)**2 * d phi);
\end{verbatim}
Before we repeat the calculation of Maxwell's equation, we again make use
of the Reduce switches {\tt exp} and {\tt gcd}:
\begin{verbatim}
    off exp$ on gcd$
    farad2  := d pot1;
    maxhom3 := d farad2;
    excit2  := (1/mu) * # farad2;
    maxinh3 := d excit2;
\end{verbatim}
The expression for {\tt maxinh3} looks rather complicated
--- if we require it to vanish, we get a system of second order PDEs in two
variables. We try solve it by inspection; in a first step first let us specify
the $g$-factor. There are two values for $g$ that seem plausible: $g = 1$
corresponds to a small ring current as considered in Sec.\ref{Dipole}, and
$g = 2$ is the $g$-factor of an electron.
As {\tt maxinh3} contains $(g - 2)$ several times as a factor, the choice
\begin{verbatim}
    g := 2$
\end{verbatim}
seems promising. Indeed, if we compute the inhomogeneous Maxwell equation
again
\begin{verbatim}
    maxinh3 := d excit2;
\end{verbatim}
the coefficient of $o^t\wedge o^\rho \wedge o^\theta$ turns out to
be quite compact. If it is to vanish, we have to solve
\beq
   \frac{2\, q\, a}{{\tt rr}^\frac{3}{2}\mu\, \rho\, f^2}
        \left(\cos \theta \pq{f}{\theta} - \rho\sin \theta \pq{f}{\rho}\right)
        = 0.
\eeq
Solving this equation by hand, fixes the function $f$ up to a constant
(which we can absorb into the charge $q$). We substitute the solution and
check the inhomogeneous field equation:
\begin{verbatim}
    f := 1 + (a/rho * cos(theta))**2$
    maxinh3;
\end{verbatim}
Note that the inhomogeneous Maxwell equation is satisfied irrespective of the
explicit form of the functions {\tt rr} and {\tt delsqrt}.\par

With this result, the final form of the potential can be checked:
\begin{verbatim}
    pot1 := pot1;
\end{verbatim}
\beq
    {\tt pot1} := - \frac{q\, \rho\, {\tt rr}}
                     {(a^2\cos^2\theta +\rho^2)\,{\tt delsqrt}}\; o^t.
\eeq

What remains to be found is the solution of the Einstein-Maxwell equation.
This is done in much the same manner as in Sec.\ref{ReissnerNordstrom} by
modifying the function {\tt delsqrt}. In addition, we now specify {\tt rr}
without modification:
\begin{verbatim}
    pform u1=0$ fdomain u1=u1(r,theta)$
    delsqrt := sqrt(rho**2 + a**2 - 2*m*rho + u1)$
    rr      := sqrt(rho**2 + (a * cos(theta))**2)$
\end{verbatim}
The field equation must determine the new function {\tt u1}. Thus we compute
the left hand side (the Einstein 3--form)
\begin{verbatim}
    riemannconx chris1$ chris1(a,b) := chris1(b,a)$
    pform curv2(a,b)=2, einstein3(a)=3$
    index_symmetries curv2(a,b):antisymmetric$
    curv2(a,b)    := d chris1(a,b) + chris1(-c,b) ^ chris1(a,c);
    einstein3(-a) := (1/2) * curv2(b,c) ^ #(o(-a)^o(-b)^o(-c));
\end{verbatim}
and the right hand side (the electromagnetic energy--momentum 3--form)
\begin{verbatim}
    pform lmax4=4, maxenergy3(a)=3$
    lmax4 := -(1/2) * farad2 ^ excit2;
    maxenergy3(-a) := e(-a) _| lmax4 + (e(-a) _| farad2)^excit2;
\end{verbatim}
of the field equation. These yield a differential equation for the function
{\tt u1}:
\begin{verbatim}
    pform ode3(a)=3$
    ode3(a) := einstein3(a) + kappa * maxenergy3(a);
\end{verbatim}
This system can be solved most easily by inspection of the $\varphi$--component
of {\tt ode3}:
\beq
   ode3^\varphi = -\frac{a \cos \theta}{2\, {\tt delsqrt}}\pq {u_1}{\theta}\,
               o^\rho \wedge o^\theta \wedge o^\varphi
             + \frac{\kappa q^2 - 2\mu\, u_1}{2\, {\tt rr}^4}\,
               o^t \wedge o^\rho \wedge o^\theta
\eeq
The first term demands that $u_1$ must not dependent on the coordinate
$\theta$. The second term vanishes if we substitute
\begin{verbatim}
    u1 := kappa / (2*mu) * q**2$
\end{verbatim}
Finally, we display the explicit form that {\tt delsqrt} evaluates to:
\begin{verbatim}
    delsqrt := delsqrt;
\end{verbatim}
\beq
    {\tt delsqrt} := \sqrt{\rho^2 - 2 m\rho
            + \frac{\kappa q^2}{2\,\mu} + a^2}
\eeq
Again, although we modified the spacetime geometry via
the function {\tt delsqrt}, we do not worry about the inhomogeneous Maxwell
equation because it has been found to be satisfied irrespective of the
explicit form of the functions contained in the coframe.\par

This completes our deduction of the Kerr-Newman solution from the Kerr solution
and the Einstein-Maxwell equations. This derivation is certainly  not very
original. However, the charged version of the Kerr solution with dynamic
torsion\cite{Baekler88a} was found by this type of technique.
\section*{Acknowledgment}
We are grateful to David Hartley for helpful comments.
\section*{Appendix: Maxwell program}
\begin{verbatim}
    % file mustermax.exi, 1995-02-15

    load_package excalc$

    % the Maxwell program is preceded by coframe and frame commands:
    % specify the coframe by filling in the exact form of the
    % appropriate coframe with your favorite editor, specify possibly
    % unknown functions (zero-forms) psi etc. and their domains

    pform psi=0$
    fdomain psi=psi(r)$

    coframe o(0)     =                  d t,
            o(1)     =                  d r,
            o(2)     = r *              d theta,
            o(3)     = r * sin(theta) * d phi
       with signature (-1,1,1,1)$
    frame e$

    % here begins the Maxwell program proper: unknown functions aa0,
    % aa1, etc. are prescribed for the components of the potential
    % 1-form; field strength farad2, excitation excit2, and the left
    % hand sides of the Maxwell equations are defined

    pform {aa0,aa1,aa2,aa3}=0, pot1=1, {farad2,excit2}=2,
          {maxhom3,maxinh3}=3$

    pot1    := aa0*o(0) + aa1*o(1) + aa2*o(2) + aa3*o(3)$
    farad2  := d pot1;
    maxhom3 := d farad2;
    excit2  := (1/mu) * # farad2;
    maxinh3 := d excit2;

    % Maxwell Lagrangian and energy-momentum current are assigned

    pform lmax4=4, maxenergy3(a)=3$

    lmax4 := -(1/2)*farad2^excit2;
    maxenergy3(-a) := e(-a) _|lmax4 + (e(-a) _|farad2)^excit2;

    end;

    % Be sure that you use a blank before the interior product sign!
\end{verbatim}

\end{document}